%% file: main.tex
\documentclass[final,num-names,3p,times]{elsarticle}

\usepackage{graphicx}
\usepackage{amsmath, amssymb, amsthm}
\usepackage{subfigure}
\usepackage{multirow}
\usepackage{xfrac}

\biboptions{sort&compress}

\journal{???}

\newtheorem{theorem}{Theorem}
\newtheorem*{nfl}{Normal form lemma}
\newtheorem{prop}{Proposition}
\newtheorem{cor}{Corollary}

\let\eps\varepsilon
\def\T{\hbox{\small T}}

\def\sim{\mathop{\mathrm{sim}}}

\def\P{\hbox{P}}
\def\NP{\hbox{NP}}
\def\npc{\hbox{NP-hard}}
\def\match{\hbox{\sc Matching}}
\def\cmatch{\hbox{\sc Cubic-Matching}}
\def\subcmatch{\hbox{\sc Subcubic-Matching}}
\def\median{\hbox{\sc Median}}
\def\bpmedian{\hbox{\sc Breakpoint-Median}}
\def\maxcut{\hbox{\sc Max-Cut}}
\def\cmaxcut{\hbox{\sc Cubic-Max-Cut}}
\def\quartet{\hbox{\sc Breakpoint-Quartet}}

\def\halving{\hbox{\sc Halving}}
\def\ggh{\hbox{\sc Guided-Genome-Halving}}
\def\phylo{\hbox{\sc Small-Phylogeny}}
\def\PHYLO{\hbox{\sc Large-Phylogeny}}
\def\G{{\cal G}}
\newcommand{\dotcup}{\mathbin{\mathaccent\cdot\cup}}

\newsavebox{\tempbox}

\begin{document}

\begin{frontmatter}
\title{On the complexity of rearrangement problems under the breakpoint distance}
\author{Jakub Kov\'a\v c}
\ead{kuko@ksp.sk}
\address{Department of Computer Science,
           Comenius University,
           Mlynsk\'a Dolina,\\842~48 Bratislava, Slovakia}

\begin{abstract}
We study complexity of rearrangement problems in the generalized breakpoint model
and settle several open questions. The model was introduced by Tannier et al.\ (2009)
who showed that the median problem is solvable in polynomial time in the multichromosomal
circular and mixed breakpoint models. This is intriguing, since in most other rearrangement
models (DCJ, reversal, unichromosomal or multilinear breakpoint models), the problem is NP-hard.
The complexity of the small or even the large phylogeny problem under the breakpoint
distance remained an open problem.

We improve the algorithm for the median problem and show that it is
equivalent to the problem of finding maximum cardinality non-bipartite
matching (under linear reduction). On the other hand, we prove that
the more general small phylogeny problem is NP-hard. Surprisingly,
we show that it is already NP-hard (or even APX-hard) for 4~species
(a quartet phylogeny). In other words, while finding an ancestor for 3 species
is easy, finding two ancestors for 4 species is already hard.

We also show that, in the unichromosomal and the multilinear breakpoint model,
the halving problem is NP-hard, thus refuting the conjecture of Tannier et al.
Interestingly, this is the first problem which is harder in the breakpoint
model than in the DCJ or reversal models.
\end{abstract}
\end{frontmatter}

\input intro
\input prelims
\input halving
\input median
\input phylo
\input conclusion

\subsubsection*{Acknowledgements.}
The autor would like to thank Bro\v na Brejov\' a for many constructive comments.
The research of Jakub Kov\'a\v c is supported by Marie Curie Fellowship IRG-224885 to Dr.\ Tom\'a\v s Vina\v r,
Comenius University grant UK/121/2011, and by National Scholarship Programme (SAIA), Slovak Republic.
A preliminary version of this paper was presented on the Ninth Annual RECOMB
Satellite Workshop on Comparative Genomics (RECOMB-CG 2011).

\bibliographystyle{elsarticle-num-names}
\bibliography{main}
\end{document}

%% file: intro.tex
\section{Introduction}

While point mutations change the genomic \emph{sequence} of species throughout the evolution,
there are also large scale rearrangement mutations, such as inversions or translocations, which affect
the \emph{order} of genes in a genome.
The gene order data can be used for inferring phylogenetic relationships
and for reconstructing phylogenies \citep{moret05}.
A related problem is the reconstruction of ancestral gene orders, which is key to understanding
the underlying evolutionary processes.


The simplest model for studying gene orders is the \emph{breakpoint model} introduced by \citet{sankoff98}.
When two genes (or conserved segments or markers) are adjacent in one genome,
but not in the other, we call this position a breakpoint. We can then define
the breakpoint distance simply by counting the number of breakpoints.

\citet{sankoff98} tried to reconstruct the ancestral gene orders, given a phylogenetic tree 
and gene orders of the extant species, based on the parsimony criterion, i.e., by minimizing
the sum of distances along the branches of the tree. 
This is known as the \emph{small phylogeny} problem\footnote{as opposed to the large phylogeny problem,
where the phylogenetic tree is not given and is part of the solution}.
Unfortunatelly, the problem is NP-hard already when we have three species -- an important
special case known as the \emph{median} problem. In fact, the median problem turns out to be
NP-hard for almost all rearrangement distances 
(breakpoint \citep{peer,bryant,tannier}, reversal \citep{caprara}, and DCJ \citep{tannier}).

One notable exception is the general breakpoint model. \citet{tannier} observed that
if we drop the condition that genomes are unichromosomal and that all chromosomes are linear,
we get a simple model where the median problem is solvable in polynomial time.
Even though this model is not very biologically plausible and more realistic models exist,
the breakpoint model may still be useful for upper and lower bounds, and solutions in this model
may serve as good starting points for the more elaborate and complicated models.

In this paper, we complete the work started by \citet{tannier} on the breakpoint model.
We study several rearrangement problems in different variants of the breakpoint model
and settle their computational complexity.

\subsection{Previous results and our contribution}

There are several variants of the breakpoint model depending on what karyotypes do we allow.
In the \emph{unichromosomal} (linear or circular) model, the genome may only consist of one chromosome.
In the \emph{multilinear model}, the genome may consist of multiple linear chromosomes and
finally, the mixed model allows for any number of linear and circular chromosomes
(even though this is not biologically plausible).

For the unichromosomal model, \citet{peer} and \citet{bryant} showed that
the median problem is NP-hard. This result was extended to the multilinear model
by \citet{tannier}, and \citet{zheng} showed the NP-hardness for a related problem
called guided halving (see Preliminaries).

Curiously, the ordinary halving problem was not studied before in the breakpoint model,
and \citet{tannier} also leave it open. Moreover, they conjecture that the problem
is polynomially solvable -- this might perhaps be attributed to the fact that the halving
problem is polynomially solvable in far more complicated models such as reversal/translocation (RT)
\citep{elmabrouk} or double cut and join (DCJ) \citep{alekseyev,mixtacki,warren,kovac}.
Nevertheless, we refute this conjecture (unless $\P=\NP$) by proving that the halving problem
is NP-complete in the unichromosomal and multilinear models.

\medskip

Our main contribution is, however, our work in the general (mixed) model.
\citet{tannier} introduced this model and showed that median, halving,
and guided halving problems are solvable in polynomial time.

Two open questions remained in the work of \citet{tannier}. These are also articulated in
the monograph by \citet{bible}:
\begin{quote}
1. The best time complexity for the median and guided halving problems under the
breakpoint distance on multichromosomal genomes (with circular chromosomes allowed)
is $O(n^3)$, using a reduction to the maximum weight perfect matching problem.
It is an open problem to devise an ad-hoc algorithm with better complexity.
\end{quote}
\begin{quote}
2. The small parsimony problem and large parsimony problem under the breakpoint distance
is open regarding multichromosomal signed genomes when linear and circular chromosomes are allowed.
\end{quote}
We resolve the first question in a positive way by showing a more efficient algorithm running
in $O(n\sqrt n)$ time. This is by reduction to the maximum cardinality matching problem.
Moreover, we show that maximum cardinality matching can be reduced back to the breakpoint
median (by a linear reduction) and so the two problems have essentially the same complexity.
The same technique also improves the algorithms for halving and guided halving.

The second question is resolved in a negative way. One could expect that the large parsimony
problem is NP-hard for this model, since it is NP-hard even for the Hamming distance on binary strings
\citep{foulds}. However, surprisingly, for the breakpoint distance (unlike the Hamming distance),
the small phylogeny is NP-hard, and it is NP-hard even for 4 species, i.e., a quartet phylogeny.
In other words, while the small phylogeny problem is easy for 3 species, it is hard already for 4 species.


The previous work and our new results are summarized in Table~\ref{tab}.

\begin{table*}[h]
\centering
\caption{Our new results in context of the previously known results.} \label{tab}
\medskip
\begin{tabular}{c@{\qquad}cccc}
\sc Breakpoint Model           & \sc Median      & \sc Halving       & \sc Guided Halving & \sc Small Phylogeny \\
\noalign{\smallskip\hrule\smallskip}
\it unichromosomal             & 
\multirow {2}*{\npc\ \citep{peer,bryant}} & 
\multirow {2}*{\npc\ [new]}               & 
\multirow {2}*{\npc\ \citep{zheng}}       & 
\multirow {2}*{\npc\ [trivially]} \\        
\it (linear or circular) \\
\noalign{\medskip}
\it multilinear &
\npc\ \citep{tannier} & 
\npc\ [new]           & 
\npc\ \citep{zheng}   & 
\npc\ [trivially] \\    
\noalign{\medskip}
\it multichromosomal &
$O(n^3)$ \citep{tannier}, & 
$O(n^3)$ \citep{tannier}, & 
$O(n^3)$ \citep{tannier}, & 
\multirow {2}*{\npc\ [new]} \\                                  
\it (circular or mixed) &
$O(n\sqrt n)$ [new] &
$O(n)$ [new]       &
$O(n\sqrt n)$ [new] 
\end{tabular}
\end{table*}

\subsection{Road map}

In the next section, we define the different variants of the breakpoint model and state the rearrangement
problems.
In Section 3, we refute the conjecture of \citet{tannier} and prove that the halving problem is NP-hard
for the unichromosomal and multilinear breakpoint model.
In the following two sections, we study the general breakpoint model. In Section 4, we look at the median
problem: we improve upon the algorithm of \citet{tannier} and show that it is equivalent to the 
maximum matching problem. The hardness of the small phylogeny problem is studied in Section 5
and we conclude in Section 6.

%% file: prelims.tex
\section{Preliminaries}

\subsection{Genome models and the breakpoint distance}

We assume that all the studied genomes have the same gene content, and we denote this set of genes by $\G$.
We also assume that each gene $g\in\G$ is an oriented segment of DNA having two ends -- a \emph{head} and a \emph{tail}.
These two ends are called \emph{extremities} and are denoted $g_h$ and $g_t$, respectively.  
Let us first describe the circular models which are easier to work with.
We then extend our definitions to account for linear chromosomes.

We represent genome $\pi$ by a set of edges: An edge between extremities $x$ and $y$, called
\emph{adjacency}, indicates that $x$ and $y$ are adjacent in the genome.
Note that in circular genomes, every extremity is adjacent to exactly one other extremity,
so we can \emph{identify genomes with perfect matchings} over the set of extremities.

\begin{figure*}[!hb]
  \centering
  \subfigure[The order of genes in a genome. Each arrow corresponds to a single gene with known orientation.]{
    \hspace{2em}\includegraphics[scale=0.9]{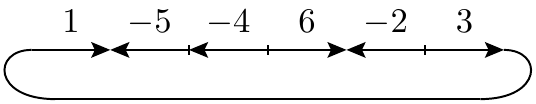}\hspace{2em}\label{fig:ge}
  }\qquad
  \subfigure[Representation of the genome on the left by a perfect matching.
  The green edges are the adjacencies of $\pi$, the gray edges form the base matching $B$.
  The Hamiltonian cycle $\pi\dotcup B$ corresponds to the single chromosome.]{
    \hspace{9em}\includegraphics[scale=1]{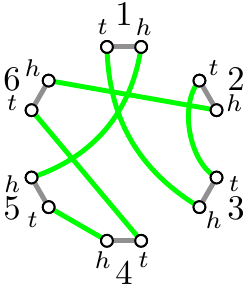}\hspace{9em}\label{fig:grb}
  }
  \caption{Example of a circular genome $\pi$ and its representation by a perfect matching.}\label{fig:example1}
\end{figure*}

Let us define an auxiliary \emph{base matching} $B = \{ g_hg_t : g\in\G \}$ where each edge connects the two ends of some gene.
Then all vertices have degree 2 in the union\footnote{technically, this is a disjoint or multiset union; we allow parallel
edges forming 2-cycles} $\pi\dotcup B$, and $\pi\dotcup B$ decomposes into a set of cycles, which naturally correspond
to the circular chromosomes of our genome (see Fig.~\ref{fig:example1}).

In the \emph{general} (multichromosomal circular) model, genomes can have multiple circular chromosomes
and any perfect matching $\pi$ corresponds to a genome.
In the unichromosomal \emph{circular} model, we require that the genome only consists of a single chromosome,
so $\pi\dotcup B$ is a Hamiltonian cycle as in Fig.~\ref{fig:example1}. Such a matching $\pi$ is sometimes called a Hamiltonian matching.

\medskip

Let $\pi_1$ and $\pi_2$ be two genomes -- two perfect matchings. Then the breakpoint distance between $\pi_1$
and $\pi_2$ is defined as
$$d(\pi_1,\pi_2) = n - \sim(\pi_1,\pi_2),$$
where $n$ is the number of genes and $\sim(\pi_1,\pi_2)$ is the number of common adjacencies.
The breakpoint distance satisfies all the properties of a metric and is used in the literature,
however, we find it easier to work directly with the similarity measure $\sim(\pi_1,\pi_2)$.

\medskip
\goodbreak

\begin{figure*}[!ht]
  \centering
  \subfigure[A genome with 2 linear and 1 circular chromosome. Such genomes are not found in nature,
  however, the model is motivated by tractability of the rearrangement problems such as median.]{
    \hspace{2em}\includegraphics[scale=0.9]{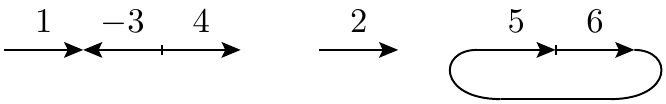}\hspace{2em}\label{fig:ge2}
  }\qquad
  \subfigure[The same genome represented as a set of adjacencies (green matching).
  Gray edges form the base matching $B$. Components of $\pi\dotcup B$ are paths and cycles,
  corresponding to linear and circular chromosomes, respectively.]{
    \hspace{6em}\includegraphics[scale=1]{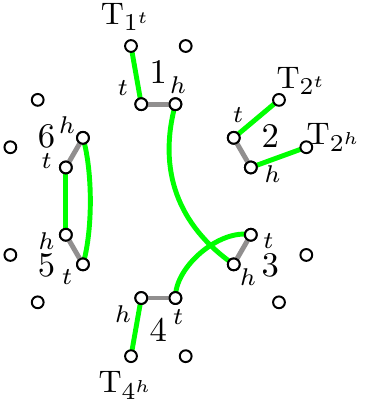}\hspace{6em}\label{fig:gr2}
  }
  \caption{Example of a mixed genome $\pi$ and its representation.}\label{fig:example2}
\end{figure*}

To represent linear chromosomes, we add a vertex $\T_x$ for each extremity $x$.
These vertices are called \emph{telomeres} and a \emph{telomeric adjacency} $x\,\T_x$
indicates that $x$ is an end of a linear chromosome (see Fig.~\ref{fig:example2}).

Genomes will again correspond to matchings with a condition that $\T_x$ may only be adjacent
to $x$. If $\pi$ is such a matching, $\pi\dotcup B$ consists of cycles and paths ending in telomeres,
which correspond to circular and linear chromosomes, respectively. In the \emph{mixed} model,
any such matching $\pi$ represents a genome; in the \emph{multilinear} model, we require that
every chromosome is linear; and in the \emph{linear} model, we only allow a single linear
chromosome.

We can write the breakpoint distance again in the form $d(\pi_1,\pi_2) = n - \sim(\pi_1,\pi_2)$,
where this time, $\sim(\pi_1,\pi_2)$ is the number of common adjacencies plus \emph{half} the number
of common telomeric adjacencies (as introduced by \citet{tannier}).

\subsection{Duplicated genomes}

We will also work with \emph{duplicated} genomes that underwent a whole genome duplication and have exactly
two copies of each gene. For each gene $g$, let us label the first copy $g^1$ and the second copy $g^2$.
Then we can represent a duplicated genome by an ordinary genome $\delta$ over the gene set $\{ g^1, g^2 : g\in \G\}$.
However, note that the labels were introduced arbitrarily
and we consider two genomes that differ only in the subscripts of some genes as equivalent.
A duplicated genome actually corresponds to the equivalence class $[\delta]$.

We can define the breakpoint distance (similarity) between two duplicated genomes $[\gamma]$ and $[\delta]$
as the minimum distance (maximum similarity) between ordinary genomes $\gamma'\in[\gamma]$ and $\delta'\in[\delta]$.
In fact, we can fix one $\gamma'\in[\gamma]$ and take the minimum (maximum) over $\delta'\in[\delta]$.

Let us write $\theta = \pi\oplus\pi$ for a \emph{perfectly} duplicated genome -- the result of a whole
genome duplication. For each linear chromosome in $\pi$, $\theta$~contains two copies of the chromosome
and for each circular chromosome in $\pi$, $\theta$~contains either two copies of the chromosome
or one chromosome consisting of the two copies consecutively.
The distance between an ordinary genome $\pi$ and a duplicated genome $[\delta]$, also called \emph{double
distance} and denoted $dd(\pi,\delta)$, is then the distance between $\pi\oplus\pi$ and $[\delta]$.

We say that $\pi$ and $[\delta]$ have adjacency $xy$ in common, if $x,y$ are adjacent in $\pi$
and $x^i,y^j$ are adjacent in $\delta$ for some $i$ and $j$. We say that they have the adjacency $xy$
twice in common, if either $x^1y^1$ and $x^2y^2$, or $x^1y^2$ and $x^2y^1$ are adjacent in $\delta$.
\citet{tannier} showed that the double distance $dd(\pi,\delta)$ can be computed simply as 
$dd(\pi,\delta) = 2n - \sim(\pi,\delta)$, where $\sim(\pi,\delta)$ is the number of adjacencies in common
plus half the number of telomeric adjacencies in common (adjacencies twice in common are counted as 2).

\subsection{Rearrangement problems}

Once we have a genome model and a distance measure, we can define the problems of interest.
In general, the focus of our study are problems related to reconstruction of ancestral genomes
under the parsimony principle.

Assume that we have two genomes $\pi_1$ and $\pi_2$, and we would like to reconstruct their common
ancestor $\alpha$. Using a third, outgroup genome $\pi_3$, we can formulate the task as the \median\
problem: Given $\pi_1$, $\pi_2$, and $\pi_3$, find genome~$\alpha$ (called \emph{median}) that minimizes the total distance
from $\pi_1$, $\pi_2$, and $\pi_3$. In the \bpmedian\ problem, we are minimizing the breakpoint distance,
which is the same as \emph{maximizing} the median score
$S(\alpha) = \sim(\alpha,\pi_1) + \sim(\alpha,\pi_2) + \sim(\alpha,\pi_3)$.
Note that the genome model imposes further constraints on the solution -- the number and type of chromosomes.

We can generalize the median problem to the median of $k$ genomes problem, where given genomes $\pi_1,\ldots,\pi_k$,
we should find genome $\alpha$ that maximizes the score $S(\alpha)=\sum_i \sim(\alpha,\pi_i)$.
However, even more important generalization is the \phylo\ problem, where we  
are given a phylogenetic tree and gene orders of the extant species (leaves of the tree).
The task is to reconstruct all the ancestral genomes, i.e., to find gene orders for each internal vertex,
while minimizing the sum of breakpoint distances along the edges of the phylogenetic tree.
(This is the same as maximizing the sum of similarities along the edges.)
The \median\ problem is a special case of the \phylo\ problem with just 3 species. On the other hand,
median solvers are widely used in practice in the Steinerization heuristic to reconstruct the ancestors
in \phylo: Starting with some initial ancestral genomes, we repeatedly replace genomes by medians of
the neighbouring genomes in the phylogeny, until we converge to some local optimum.
Therefore, having a model where the \median\ problem is efficiently solvable might be of practical significance.

Another classical problem in genome rearrangements is the \halving\ problem. Imagine a genome $\pi$ that underwent
a whole genome duplication. The perfectly duplicated genome $\theta = \pi\oplus\pi$ was then rearranged
to its present-day form $\gamma$. In the \halving\ problem, we would like to reconstruct
the pre-duplication ancestor $\pi$ given the present-day genome $\gamma$. More precisely, we would like to
find an ordinary genome $\alpha$ that minimizes the double distance from $\gamma$.

The \halving\ problem has usually many equivalent solutions. For better results, we can use an
ordinary outgroup genome $\rho$ (such that the speciation happened before the whole genome duplication) 
and search for genome 
$\alpha$ that minimizes the sum $dd(\alpha,\gamma)+d(\alpha,\rho)$. This is called the \ggh\ problem.

%% file: halving.tex
\section{The halving problem}

\citet{bryant} showed that the median problem is NP-hard in the circular breakpoint model
by reduction from the {\sc Directed-Hamiltonian-Cycle} problem. The halving problem was not studied
previously in the breakpoint model, but we show that it suffers the same ``Hamiltonian" curse
as the median problem -- in order to find the ancestor, we would in fact have to find a Hamiltonian
cycle. Our proof is even simpler than that of \citet{bryant}.

As the halving problem is polynomially solvable in more realistic models such as the RT model
\citep{elmabrouk} or the DCJ model \citep{alekseyev,mixtacki,warren,kovac}, 
the halving problem under the breakpoint distance will remain a mere curiosity:
It is the first problem which is easier in the DCJ or even in the RT model than in the breakpoint model.
Furthermore, it is the only known case where halving is NP-hard, while the double distance
is computable in polynomial time (e.g., in the DCJ model, the opposite is true -- halving is easy,
while the double distance is NP-hard \citep{tannier}).

\begin{theorem}
Halving problem is NP-hard in the circular, linear, and multilinear breakpoint models.
\end{theorem}

\proof
The proof is by reduction from the {\sc Directed-Hamiltonian-Cycle} problem.
\citet{plesnik} proved that this problem is still NP-hard for graphs with
maximum degree 2 and the construction implies the problem is also NP-hard if
all in-degrees and out-degrees are equal to 2. Note that such graphs
have an Eulerian cycle.

Let $G=(V,E)$ be such a directed graph; the corresponding doubled genome $\delta$
will have two copies of a gene for each vertex in $G$ and an Eulerian cycle
in $G$ traversing each vertex twice will be the order of genes in $\delta$.
More precisely, let $G'=(V',E')$, where $V' = \{ x^1_h, x^1_t, x^2_h, x^2_t : x \in V \}$
and the edges in $E'$ are defined as follows: traverse the Eulerian walk and for each edge $xy\in E$,
include an edge $x^i_h y^j_t$ in $E'$, where $i$ and $j$ is 1, if we are visiting the vertex
for the first time and 2, if we are visiting the vertex for the second time.
Note that all edges go from head to tail, $E'$ is a perfect matching,
and $G'$ defines the doubled genome $\delta$ consisting of a single circular chromosome.

Let $\alpha$ be a circular genome, a solution to the halving problem.
Note that $\delta$ has no double adjacencies, so $\alpha$ can have at most
$n$ adjacencies in common (none twice in common).
This maximum can be attained if and only if all the adjacencies in $\alpha$
are of the form $x_hy_t$ (from head to tail) and for each such adjacency,
$x^i_hy^j_t$ is an adjacency in $\delta$ for some $i,j$. This is if and only
if $xy\in E$. So by contracting the base matching (each head and tail of
a gene into a single vertex) and orienting the edges (from head to tail),
we get a directed Hamiltonian cycle in $G$.


For the linear and multilinear models, remove one edge $xy$ from $G$ and consider
the problem of deciding whether $G$ contains a directed Hamiltonian path. This problem is still
NP-hard and can be reduced to the halving problem in the linear models:
$G$ now has an Eulerian path starting in $y$ 
and ending in $x$. We replace the last adjacency $x^2_h y^1_t$ in $\delta$
(corresponding to the removed edge) by two telomeric adjacencies $x^2_h\T_{x^2_h}$ and $y^1_t\T_{y^1_t}$
to get a linear genome.
If $\alpha$ is a linear or multilinear solution to the halving problem, it can reach the maximum
similarity if and only if all its adjacencies (including the telomeric adjacencies) are in common
with $\delta$ and this is if and only if contraction of $\alpha$ is a directed Hamiltonian path in $G$.
\qed

%% file: median.tex
\section{Median and halving problems in the general model}

From now on, we will study the \emph{general} breakpoint model,
i.e., the multichromosomal circular model where genomes are
perfect matchings. We will also note how to extend the results
to the mixed model and use the developed techniques for the
halving and guided halving problem.

\subsection{Breakpoint median}

\citet{tannier} noticed that finding a breakpoint median can
be reduced to finding a maximum weight perfect matching.
This can be done in $O(n^3)$ time by algorithm of \citet{gabow73}
and \citet{lawler76}. An open problem from \citet{tannier}
and \citet{bible} asks, whether this can be improved.
We answer this question affirmatively by showing an $O(n\sqrt{n})$
algorithm.

The solution by \citet{tannier} (if we rephrase it using the similarity measure
instead of the breakpoint distance) was to create a complete weighted graph $G$
where vertices are extremities and weight $w(xy)$ of edge $xy$ is the number of
genomes which contain the adjacency $xy$. Any perfect matching $\alpha$
corresponds to some genome and the weight of the matching is equal to its median
score $S(\alpha)$.

Notice that instead of finding a maximum weight \emph{perfect}
matching, we can remove all the zero-weight edges from $G$ and find
an ordinary (not necessarily perfect) matching.
We can then complete the genome by joining the free vertices arbitrarily.
Since the number of edges in $G$ is now linear, maximum weight matching can be found in
$O(n^2\log n)$ time by algorithm of \citet{gabow90}
or even in $\tilde O(n\sqrt{n})$ time by the state of art algorithm of \citet{gabow91}
using the fact that the weights are small integers.
More generally and more precisely:

\begin{theorem}\label{thm:med-k}
The \bpmedian\ problem for $k$ genomes can be solved 
in $O(kn\sqrt{n}\cdot\log(kn)\sqrt{\mathop{\alpha}(kn,n)\log n})$ 
time in the general model. (Here, $\alpha(m,n)$ is the inverse Ackermann function.)
\end{theorem}

We further improve the algorithm for the most important special case, $k=3$:
Notice that when $xy$ is an edge with weight 3, there is no other
edge incident to $x$ or $y$. Therefore, $xy$ must belong to the maximum weight
matching. Moreover, if $xy$ has weight 2, there is a maximum weight matching
which contains $xy$. Suppose to the contrary that $xu$ and $yv$ were matched in $\alpha$ instead.
Then $w(xu)$ and $w(yv)$ is at most 1 and by exchanging these edges for $xy$ and $uv$
with weights $w(xy)=2$ and $w(uv)\geq0$ we get a matching with the same or even higher weight.

Thus, we can include all edges of weight 2 and 3 in the matching and remove the matched vertices
together with their incident edges. The remaining graph has only unit edge weights,
so it suffices to find maximum \emph{cardinality} matching. This can be done in $O(m\sqrt{n})$ time
by algorithm of \citet{micali80}. Thus, we have

\begin{theorem}\label{thm:med-upper}
The \bpmedian\ problem for 3 genomes 
can be solved in $O(n\sqrt n)$ time (in the general model).
\end{theorem}

One might still wonder whether there is an even better algorithm for the median problem, which
perhaps avoids the computation of maximum matching. Alas, we show that improving upon our result
would be very hard, since it would immediately imply a better algorithm for the matching problem,
beating the result of \citet{micali80} (at least on cubic graphs), which is an open problem for more
than 30 years.

\citet{biedl} showed that the maximum matching problem is reducible to
maximum matching problem in \emph{cubic} graphs by a \emph{linear} reduction.
This means that we can transform any given graph $G$ with $m$ edges to a cubic graph
$G'$ with $O(m)$ edges such that maximum matching in $G$ can be
recovered from one in $G'$ in $O(m)$ time. Thus, any $O(f(m))$ algorithm
for maximum matching in cubic graphs implies an $O(f(m)+m)$ algorithm for
arbitrary graphs. 

We say that a reduction is \emph{strongly linear}, if it is linear
and both the number of vertices and the number of edges increase at most linearly.
Such a reduction preserves the running time $O(f(m,n))$ depending on both
the number of vertices and the number of edges.

We prove that the \bpmedian\ problem is equivalent to \match\ under linear
reduction and to \cmatch\ under strongly linear reduction. If we write
$\leq_\ell$ for linear and $\leq_{s\ell}$ for strongly linear reduction, we have
$$ \match \leq_\ell \cmatch \leq_{s\ell} \bpmedian \leq_{s\ell} \match. $$
The first reduction is by \citet{biedl} and the last one was shown in Theorem~\ref{thm:med-upper}
(in fact, a reduction to \subcmatch, where the degrees are $\leq3$, was shown -- this
is equivalent to \cmatch\ under the strongly linear reduction \citep{biedl}).
We now prove the middle reduction.

Let $G$ be a cubic graph, an instance of the \cmatch\ problem.
The difference between the \cmatch\ and \bpmedian\ problem is that
in \bpmedian, the input multigraph consists of three perfect matchings,
i.e., is edge 3-colorable. However, not all cubic graphs are edge 3-colorable (take
for example Petersen's graph).

\begin{figure*}[t]
  \centering
  \subfigure[Edge $xy$ (top) should be colored green
  (this is the only missing color at $x$) and red at the same time (this is the missing color at $y$).
  We resolve this conflict by subdividing edge $xy$ by two new vertices (bottom); we color $xu$ green,
  $vy$ red and $uv$ blue.]{
    \hspace{5em}\includegraphics[scale=1]{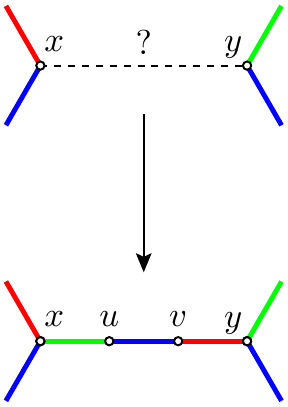}\hspace{5em}
    \label{fig:conflict}
  }\qquad
  \subfigure[In the second phase, we duplicate graph $G$ and connect the corresponding vertices with degree 2
  as shown in the figure.]{
    \includegraphics[scale=0.8]{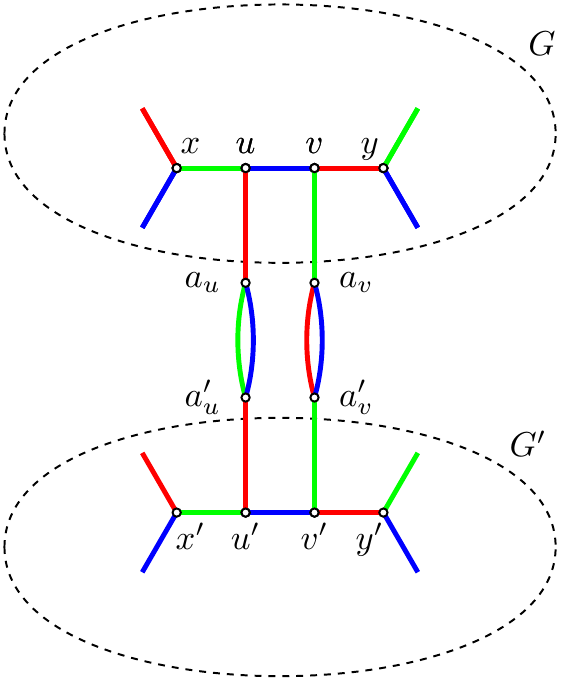}
    \label{fig:dupl}
  }
  \caption{Linear reduction of maximum matching in cubic graphs to breakpoint median problem.} 
\end{figure*}

The solution is to color edges arbitrarily and resolve conflicts as shown
in Figure~\ref{fig:conflict}. We can for example color the ends of edges
at each vertex randomly by three different colors. When both ends of an edge
are assigned the same color, we color the edge appropriatelly. When the ends
have different color, we subdivide the edge into three parts and use the
third color for the middle edge (see Figure~\ref{fig:conflict}).
Note that the size of a maximum matching in the modified graph is exactly one
more than the size in the original graph: If $xy$ is matched in the original,
$xu$ and $vy$ can be matched in the modified graph. If $xy$ is not matched,
we can still match $uv$.

Now, the modified graph is edge 3-colorable but not cubic. We remedy this by duplicating
the whole graph and connecting the corresponding vertices of low degree as shown
in Figure~\ref{fig:dupl}. As noted above, we may suppose that the auxilliary
double edges $a_ua_u'$ and $a_va_v'$ are matched, so $ua_u$, $u'a_u'$, $va_v$,
and $v'a_v'$ are not matched and given the solution for the \bpmedian\ problem,
we can recover the maximum matching of $G$ in $O(n)$ time. The reduction
is obviously linear, so we have

\begin{theorem}\label{thm:med-lower}
The \bpmedian\ problem (in the general model) has the same complexity
as finding maximum cardinality matching in cubic graphs.
\end{theorem}

\subsection{Median in the mixed model}

In the mixed model, weight of a telomeric adjacency $x\T_x$ is equal to half the number of
genomes that contain $x\T_x$. If we multiply all weights by 2, we can use the algorithm
by \citet{gabow91} for integer weights, so the result of Theorem~\ref{thm:med-k} remains
valid also in the mixed model.

For the median of 3 genomes, an $O(n\sqrt n)$ algorithm exists:
We observed that we can include all the double and tripple adjacencies in the matching.
This is also true for the double and tripple telomeric adjacencies (edges of weight 1 and $1\sfrac12$):
If $w(x\T_x)=1\sfrac12$, $x\T_x$ is a tripple adjacency and no other edge is incident to neither $x$ nor $\T_x$ in $G$.
If $w(x\T_x)=1$ but the median $\alpha$ contains adjacency $xy$ instead, then $w(xy)\leq 1$
and since $\T_x$ can only be incident to $x$, it must be unmatched (or matched by a zero-weight edge) and
so we can replace $xy$ by $x\T_x$ in $\alpha$.

The remaining graph consists of edges with unit weight and weight $\sfrac12$. Note however that all the $\sfrac12$-weight
edges are of the form $x\T_x$ and there is no other edge incident to $\T_x$. We use the doubling trick again:
we take two copies of graph $G$, and replace all pairs $x\T_x$, $x'\T_x'$
by a single edge $xx'$ of unit weight. We can then remove all the telomere vertices.
The resulting graph will have only unit weight edges and maximum matching exactly twice the size of maximum
matching in the original graph.

\subsection{Halving problems in the general model}

The same tricks can be used for the halving and the guided halving problem.
Recall that in the halving problem, given a duplicated genome $\gamma$, we
are searching for $\alpha$ that minimizes the double distance $dd(\alpha,\gamma)$
and in the guided halving problem, we are in addition given genome $\rho$ and
we are minimizing the sum $dd(\alpha,\gamma) + d(\alpha,\rho)$.

Again, we construct graph $G$, where this time, weight of edge $xy$ is
the number of adjacencies among $x^1y^1$, $x^1y^2$, $x^2y^1$, $x^2y^2$ in $\gamma$
and possibly $xy$ in $\rho$ (in case of the guided halving problem).
The rest of the solution is identical, leading to an $O(n\sqrt n)$ algorithm for the guided halving problem.
In the halving problem, the degrees of vertices in $G$ are at most 2 and after including
all the double edges in the solution, the remaining graph consists only of cycles
and the maximum matching can be found trivially in $O(n)$ time.

%% file: phylo.tex
\section{Breakpoint phylogeny}

In the \phylo\ problem, we try to reconstruct ancestral genomes given
a phylogenetic tree and gene orders of the extant species while minimizing
the sum of distances along the edges of the tree. This problem is NP-hard
for most rearrangement distances and for most models; this follows trivially
from the NP-hardness of the \median\ problem.
However, as we have seen in the previous section, this is not the case
in the general breakpoint model and the complexity of the \phylo\ problem 
remained open \citep{tannier,bible}.

In this section, we prove that the \phylo\ problem is NP-hard also in the general breakpoint model.
We show that the problem is NP-hard already for 4 species, a special case that we call the \quartet\ problem.

\begin{figure}[hp]
  \centering
  \includegraphics[scale=1]{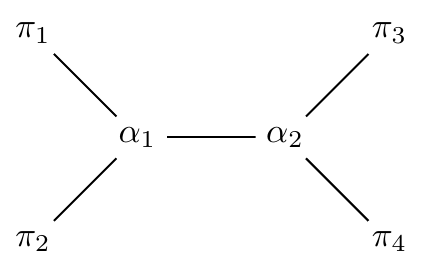}
  \caption{Quartet tree.}\label{fig:quartet}
\end{figure}

Given four genomes $\pi_1,\pi_2,\pi_3,\pi_4$, the \quartet\ problem
is to find ancestral genomes $\alpha_1,\alpha_2$ that
maximize the sum of similarities along the edges of the quartet tree
in Figure~\ref{fig:quartet}, i.e., the sum
$$ S(\alpha_1,\alpha_2) = 
   \sim(\pi_1,\alpha_1) + \sim(\pi_2,\alpha_1) + \sim(\alpha_1,\alpha_2) + \sim(\alpha_2,\pi_3) + \sim(\alpha_2,\pi_4). $$

\begin{theorem}\label{thm:quartet}
The \quartet\ problem 
is NP-hard and even APX-hard in the general breakpoint model.
\end{theorem}

The proof is inspired by the work of \citet{dees}\ who showed that the following problem is NP-hard:
Given two graphs $G_1=(V,E_1)$, $G_2=(V,E_2)$, find two perfect matchings $M_1\subseteq E_1$
and $M_2\subseteq E_2$ with the maximum overlap $M_1\cap M_2$. The problem is NP-hard even when
the components in $G_1$ and $G_2$ are just cycles.
In our proof, $\pi_1\cup\pi_2$ will correspond to $E_1$, $\pi_3\cup\pi_4$ will correspond to $E_2$,
and the unknown ancestors $\alpha_1,\alpha_2$ will correspond to the unknown perfect matchings $M_1,M_2$.

Our proof is however much more involved and there are two reasons for this: First, 
the problem formulation does not guarantee that $\alpha_1\subseteq \pi_1\cup\pi_2$
and $\alpha_2\subseteq \pi_3\cup\pi_4$. We will say that a solution $\alpha_1,\alpha_2$
that satisfies this condition is in a \emph{normal form}. The hard part of the proof is 
showing that we can transform any solution $\alpha_1,\alpha_2$ into at least as good
solution $\alpha_1',\alpha_2'$ that is in the normal form.

The second major difficulty is that we are maximizing the sum $S(\alpha_1,\alpha_2)$ instead
of just the size of the intersection. So a solution with maximum score $S(\alpha_1,\alpha_2)$
does not necessarilly maximize the term $\sim(\alpha_1,\alpha_2)$, the size of the intersection.
To overcome these difficulties, we had to modify the edge gadget from the original proof
and use a more restricted problem for the reduction.

\subsection{Overview of the proof}

The proof is by reduction from the \cmaxcut\ problem.
Given a graph $G$, the \maxcut\ problem is to find a cut of maximum size.
We may phrase this as a problem of coloring all vertices in $G$ red or green
while maximizing the number of red-green edges. 
(Partition of $V$ into the red part and the green part defines a cut
and its size is the number of edges with endpoints of different color.)
In the \cmaxcut\ problem, the instances are cubic graphs; this
variant is still NP-hard and APX-hard \citep{maxcut}.

Let $G=(V,E)$ be a given cubic graph, instance of the \cmaxcut\ problem.
We will construct genomes $\pi_1$, $\pi_2$, $\pi_3$, and $\pi_4$
such that the maximum cut in $G$ can be recovered from the solution $\alpha_1,\alpha_2$
of the \quartet\ problem in polynomial time.

\begin{figure*}[h!p]
  \centering
  \subfigure[Vertex gadget.]{
    \qquad\includegraphics[scale=0.8]{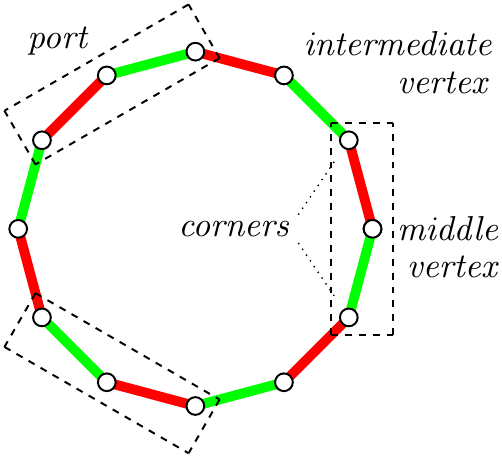}
    \label{fig:vg}
  }\qquad
  \subfigure[Edge gadget.]{
    \includegraphics[scale=0.7]{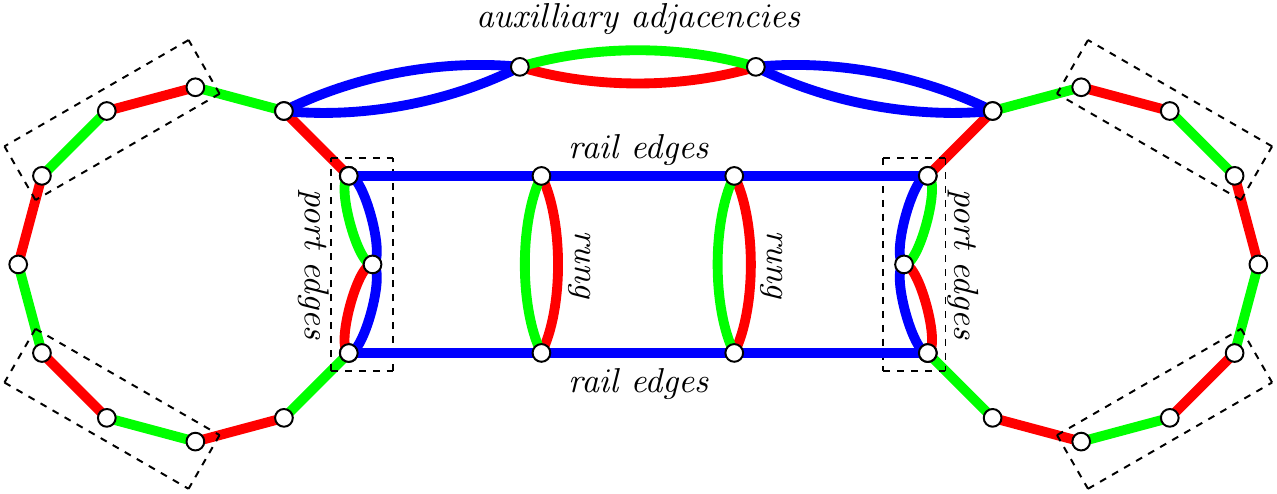}
    \label{fig:eg}
  }
  \caption{The vertex and edge gadgets used in our reduction and the terminology used for different types of
  vertices and edges. The red and green edges are the adjacencies
  of $\pi_1$ and $\pi_2$, respectively. The cycles made of blue edges can be decomposed into two matchings
  -- the adjacencies of $\pi_3$ and $\pi_4$.}
\end{figure*}

For each vertex of $G$, there will be a vertex gadget (see Figure~\ref{fig:vg})
made of adjacencies of $\pi_1$ and $\pi_2$. Let $\pi_1$ be the red matching
and $\pi_2$ the green matching. As we will prove later, we may suppose
that $\alpha_1\subseteq \pi_1\cup\pi_2$, so within each vertex gadget,
$\alpha_1$ will contain either the red edges of $\pi_1$ or the green edges of $\pi_2$.
This naturally corresponds to a red/green vertex coloring in the \cmaxcut\ problem.

The framed vertices in Figure~\ref{fig:vg} are called ``ports" -- this is where the three incident edges
are attached. For each edge of $G$, an edge gadget is constructed as shown in Figure~\ref{fig:eg}.
The blue cycles consist of two matchings -- the adjacencies of $\pi_3$ and $\pi_4$. Again, as we will prove later,
we may suppose that $\alpha_2\subseteq \pi_3\cup\pi_4$, i.e., the second ancestor consists only of the blue edges. 

For future reference, let us state here again the claims to be proved in the form of a lemma:
\begin{nfl}
Let $\pi_1,\pi_2,\pi_3,\pi_4$ be an instance of the \quartet\ problem constructed
from a \cmaxcut\ instance as described above. Then any solution $\alpha_1,\alpha_2$
can be transformed in polynomial time into a solution $\alpha_1',\alpha_2'$ such that
$S(\alpha_1',\alpha_2') \geq S(\alpha_1,\alpha_2)$ and
$$\alpha_1' \subseteq \pi_1\cup\pi_2 \hbox{~~~and~~~} \alpha_2' \subseteq \pi_3\cup\pi_4.$$
\end{nfl}

\begin{figure*}[htp]
  \subfigure[Adjacencies of the first ancestor $\alpha_1$ (red edges) agree with the adjacencies
  of $\pi_1$ at both vertex gadgets. This corresponds to coloring both vertices red in the \cmaxcut\ problem.
  Note that $\alpha_1$ and $\alpha_2$ can only have one adjacency in common.]{
    \includegraphics[scale=0.6]{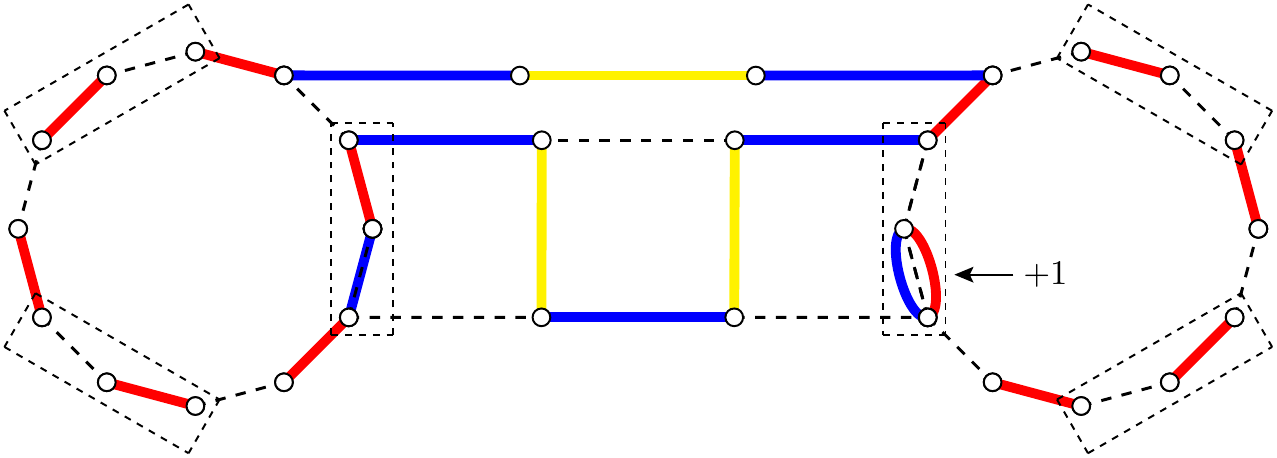}
    \label{fig:aa}
  }\hfill
  \subfigure[In the first vertex gadget, $\alpha_1$ agrees with $\pi_1$ (red edges) and in the second gadget,
  $\alpha_1$ agrees with $\pi_2$ (green edges). This corresponds to coloring the first vertex red and the second
  vertex green in the \cmaxcut\ problem. In this case, $\alpha_1$ and $\alpha_2$ have two adjacencies in common.]{
    \includegraphics[scale=0.6]{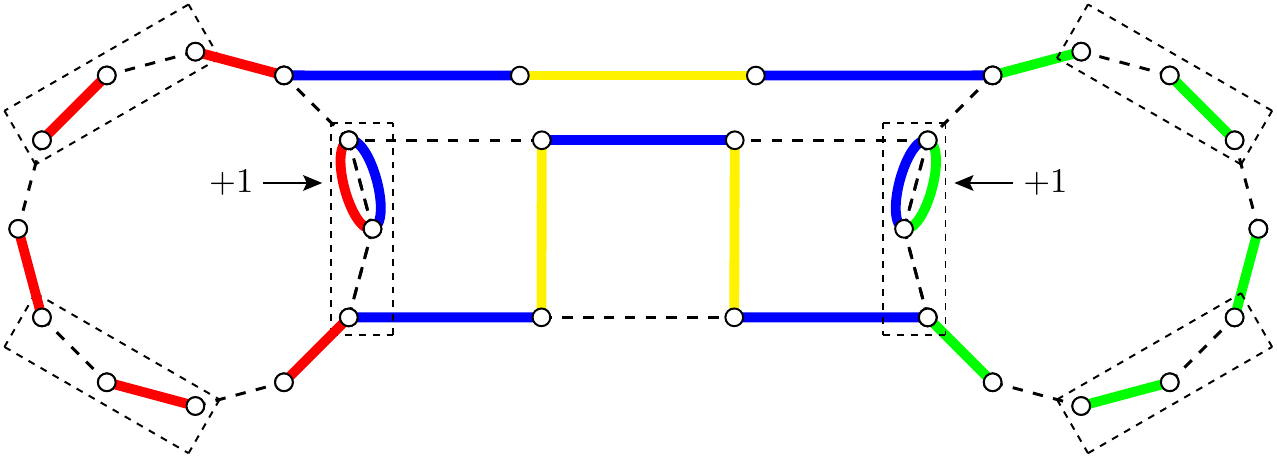}
    \label{fig:ab}
  }
  \caption{The dashed edges indicate the underlying vertex and edge gadgets, the blue edges are adjacencies of $\alpha_2$
  and the red, green, and yellow edges are adjacencies of $\alpha_1$. Here, we assume that $\alpha_1$ and $\alpha_2$ are
  in the normal form.}
\end{figure*}

Once we prove the Normal form lemma, the rest of the proof is easy: 
If $\alpha_1,\alpha_2$ is any solution in the normal form,
term $\sim(\pi_1,\alpha_1)+\sim(\pi_2,\alpha_1)$ is always the same -- we get $+6$ for each vertex gadget
and $+6$ for each edge gadget. Similarly, term $\sim(\alpha_2,\pi_3)+\sim(\alpha_2,\pi_4)$ 
is always the same -- we get $+9$ for each edge gadget. 
So the score $S(\alpha_1,\alpha_2)$ is maximized, when $\sim(\alpha_1,\alpha_2) = |\alpha_1\cap\alpha_2|$ is maximized.
Let $uv$ be an edge in our graph $G$ from the \cmaxcut\ problem; if we choose matchings of
the same color for both vertex gadgets $u$ and $v$,
then $\alpha_1$ and $\alpha_2$ can only have one edge in common within the edge gadget $uv$
(see Figure~\ref{fig:aa}).
However, if $u$ and $v$ have matchings of different color, we can set adjacencies
of $\alpha_2$ so that $\alpha_1$ and $\alpha_2$ have 2 edges in common (see Figure~\ref{fig:ab}).
When we sum up all the contributions, we get $S(\alpha_1,\alpha_2) = 20m + c$, where $m$ is
the number of edges in $G$ and $c$ is the size of the cut corresponding to the matching
$\alpha_1$, so a polynomial algorithm for \quartet\ would imply a polynomial algorithm for \cmaxcut.

For the APX-hardness, note that for any graph with $m$ edges, we can easily find a cut
of size $c \geq m/2$. Let $\alpha_1^*,\alpha_2^*$ be an optimal solution for an instance
of the \quartet\ problem and $\alpha_1,\alpha_2$ a solution such that $S(\alpha_1^*,\alpha_2^*)
\leq (1+\eps)S(\alpha_1,\alpha_2)$. Let both solutions be in the normal form and let $c^*$ and $c\geq m/2$
be the sizes of the corresponding cuts. Then $20m+c^* \leq (1+\eps)(20m+c)$ and
$c^* \leq (1+\eps)c + 20\eps m \leq (1+41\eps)c$.
So a $(1+\eps)$-approximation algorithm for the \quartet\ problem would lead to a $(1+41\eps)$-approximation
algorithm for the \cmaxcut\ problem.

It can be also proved that the phylogenetic tree $((\pi_1,\pi_2), (\pi_3,\pi_4))$ is the most parsimonous.
The alternative quartets $((\pi_1,\pi_3), (\pi_2,\pi_4))$ and $((\pi_1,\pi_4), (\pi_2,\pi_3))$ yield score
$\leq 20m$, so this result also implies the NP- and APX-hardness of the \PHYLO\ problem. It remains an
open problem whether computing the correct quartet (without reconstructing the ancestors) is hard.

\subsection{Notation, terminology, and other conventions}

We say that an adjacency $e\in\alpha_1$ is \emph{supported}, if $e\in \pi_1\cup\pi_2$. Similarly, $e\in\alpha_2$
is \emph{supported}, if $e\in\pi_3\cup\pi_4$. An adjacency that is no supported is \emph{unsupported}.
Furthermore, let $\Pi=\pi_1\cup\pi_2\cup\pi_3\cup\pi_4$ be the set of adjacencies present in at least one extant species.
We will say that an adjacency $e\in\alpha_i$ is \emph{weakly supported}, if $e\in\Pi$.

Let us name the different types of vertices (extremities) and edges (adjacencies) in the following manner:
The framed vertices in Fig.~\ref{fig:vg} are called \emph{ports} and edges from $\pi_1\cup\pi_2$ that
connect them are called \emph{port} edges. We use the same names also for other (extant or ancestral)
adjacencies which are parallel to these.

Each port consists of two outer extremities called \emph{corners} and the \emph{middle} vertex in-between.
The set of all ports, corners, and middle vertices is denoted by $P$, $C$, and $M$, respectively ($P=C\cup M$).
The set of \emph{intermediate} extremities located between ports of vertex gadgets is denoted by $I$.

The double edges and the two vertices at the top of Fig.~\ref{fig:eg} are \emph{auxilliary} -- they
just complete the matchings into perfect matchings.

Since the edge gadget without auxilliary and port edges reminds of a \emph{ladder}, we use the following
terminology (see Fig.~\ref{fig:eg}): The red-green double adjacencies are the \emph{rungs} and the blue
adjacencies are the \emph{rails} of the ladder. Again, we use the same name for parallel adjacencies.
The set of auxilliary extremities is denoted by $A$ and the set of ladder extremities is denoted by $L$.

We say that $uv$ is an $X$--$Y$-edge if $u\in X$ and $v\in Y$ ($X$ and $Y$ do not have to be disjoint);
an $X$-edge is any edge $uv$ such that $u\in X$ or $v\in X$.

In the proof of the Normal form lemma, we will gradually transform a given solution $\alpha_1,\alpha_2$
by exchanging some adjacencies in the solution for other adjacencies. The method is analogous to improving
a given matching by an augmenting path: An $\alpha_i$-\emph{alternating} cycle is a cycle where edges
belonging to $\alpha_i$ and edges not belonging to $\alpha_i$ alternate. 
We will say that $C_1,C_2$ is a \emph{non-negative pair of cycles} for the solution $\alpha_1,\alpha_2$,
if $C_i$ is an $\alpha_i$-alternating cycle and exchanging the matched and the unmatched edges
of $C_i$ in $\alpha_i$ (for $i=1,2$) does not decrease the score:
$$S(\alpha_1\oplus C_1, \alpha_2\oplus C_2) \geq S(\alpha_1,\alpha_2).$$
One of the cycles may be empty, in which case we simply say that $C_1$ or $C_2$ is a \emph{non-negative cycle}
and if the exchange in fact increases the score, we may speak of an \emph{augmenting} pair of cycles (or an
augmenting cycle).

In the figures that follow, we will draw adjacencies of $\alpha_2$ blue and adjacencies of $\alpha_1$ red, green, or yellow:
We use red and green for edges in the vertex gadgets that are in common with $\pi_1$ or $\pi_2$, respectively
(since this corresponds to choosing the red or green color in the \cmaxcut\ problem).
We use yellow for the other edges. We use straight lines for the actual adjacencies and wavy lines
for the suggested adjacencies in non-negative cycles that should be included instead.

In the proof, we will often say
\begin{quote}\it
we may suppose that the solution has property $\cal P$
\end{quote}
as a shorthand for a more precise (and longer) statement
\begin{quote}\it
Given any solution $\alpha_1,\alpha_2$, we can transform it to a solution
$\alpha_1',\alpha_2'$ with $S(\alpha_1',\alpha_2')\geq S(\alpha_1,\alpha_2)$
having propery $\cal P$ in polynomial time;
in particular, if $\alpha_1,\alpha_2$ is an optimal soluion,
$\alpha_1',\alpha_2'$ is also optimal, with property $\cal P$.
From now on, we will assume that the solution has property $\cal P$.
\end{quote}
With this terminology, we may rephrase the Normal formal lemma as follows:
\vspace{-\baselineskip}\begin{quote}\begin{nfl}
We may suppose that all adjacencies are supported.
\end{nfl}\end{quote}

\subsection{Proof of the Normal form lemma}

First, we focus on the adjacencies that the ancestors $\alpha_1$ and $\alpha_2$ 
have in common. We will show that these may be assumed to be at least weakly supported.

\begin{prop}\label{p:med}
We may suppose that all red-green double edges (auxilliary adjacencies and rungs)
are matched in $\alpha_1$ and all blue double edges (auxilliary adjacencies)
are matched in $\alpha_2$, i.e., $\pi_1\cap\pi_2 \subseteq \alpha_1$
and $\pi_3\cap\pi_4 \subseteq \alpha_2$.
\end{prop}

\begin{proof}
We can alternately replace genome $\alpha_1$ or $\alpha_2$ by the median
of its neighbors in the phylogenetic tree until we converge to a local optimum. As we have already
proved in the previous section, we may assume that a median contains all
adjacencies occuring at least twice.
\end{proof}


\begin{prop}\label{p:nom}
We may suppose that $\alpha_1$ and $\alpha_2$ do not contain unsupported $M$-edges.
In other words, we may suppose that in both $\alpha_1$ and $\alpha_2$, one of the edges
in each port is chosen.
\end{prop}
\begin{proof}
Let $x\in M$. First, consider the case that $xy_1 \in \alpha_1$ and $xy_2 \in \alpha_2$ are both unsupported. 
Let $p$ be a neighbouring corner vertex. While $xy_1$ and $xy_2$ contribute only at most $+1$ to the score (if $y_1=y_2$),
a common adjacency $xp$ would contribute $+3$. Let $pz_1$ and $pz_2$ be the actual adjacencies
in $\alpha_1$ and $\alpha_2$; either $z_1\neq z_2$, or $z_1=z_2$ and one of the adjacencies is unsupported.
Either way, these two edges contribute at most $+2$ to the score; so $xpz_1y_1x$ and $xpz_2y_2x$
is a non-negative pair of cycles and we can exchange the edges.

Similarly, if one ancestor contains a port edge $xp$ and the other one adjacencies $pz$ and unsupported $xy$,
then $xpzyx$ is a non-negative cycle.
\end{proof}


\begin{prop}\label{p:ladder}
We may suppose that all $L$-edges are weakly supported -- they are ladder edges.
\end{prop}
\begin{proof}
In $\alpha_1$, all $L$-edges are the rung edges by Proposition~\ref{p:med} and are supported. 
Consequently, contribution of any $L$-edge in $\alpha_2$ that is not even weakly supported is zero.
Let $\ell_1 x\in \alpha_2$ be such an edge. Let $\ell_1 \ell_2$ be the middle
rail edge and let $\ell_2 y$ be the adjacency in $\alpha_2$.
If $\ell_2y$ is not weakly supported, $\ell_1\ell_2yx\ell_1$ is an augmenting cycle.
Otherwise, if $\ell_2y$ is a rail edge, it contributes $+1$ to the score and $\ell_1\ell_2yx\ell_1$
is a non-negative cycle. 

The last case is that $\ell_2y$ is a rung edge contributing $+1$ to the score.
Let $\ell_3=y$, let $\ell_3\ell_4$ be the other middle rail edge, and let $\ell_4z$
be the adjacency in $\alpha_2$. Again, if $\ell_4z$ is unsupported, $\ell_1\ell_2\ell_3\ell_4zx\ell_1$ is an augmenting cycle,
otherwise it is a rail edge and the cycle is non-negative.
(We could not have stopped with the non-negative cycle $\ell_1\ell_2\ell_3x\ell_1$, since exchanging
the edges would create new unsupported $L$-edges.)

It is easy to check that with each non-negative pair of cycles, we get rid of an $L$-edge that is not
weakly supported, unless we improve the score, which may be done only $O(n)$ times. In the process,
we may introduce unsupported $C$-edges, which is okay and we will deal with them next.
\end{proof}



\begin{figure*}[th]
  \centering
  \subfigure[Case 1: $\alpha_2$ contains $m_2d$ and $cz\notin\alpha_1$.]{
    \hspace{2em}\includegraphics[scale=0.7]{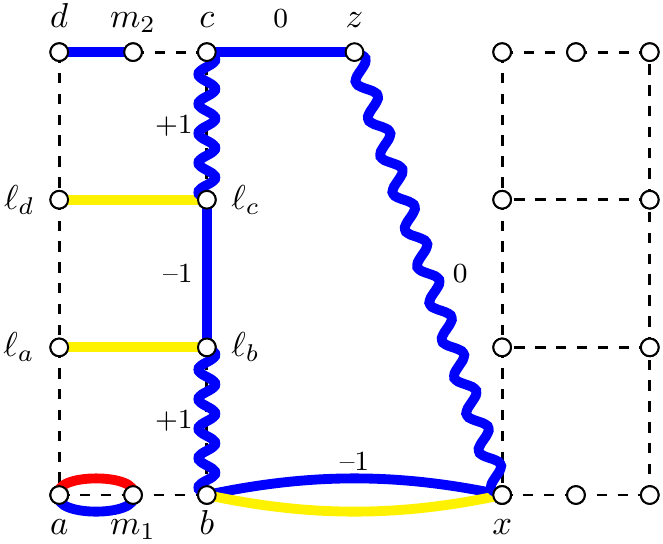}\hspace{2em}\label{fig:cc1}
  }\qquad
  \subfigure[Case 2: $\alpha_2$ contains $m_2d$ and $cz\in\alpha_1\cap\alpha_2$.]{ 
    \hspace{2em}\includegraphics[scale=0.7]{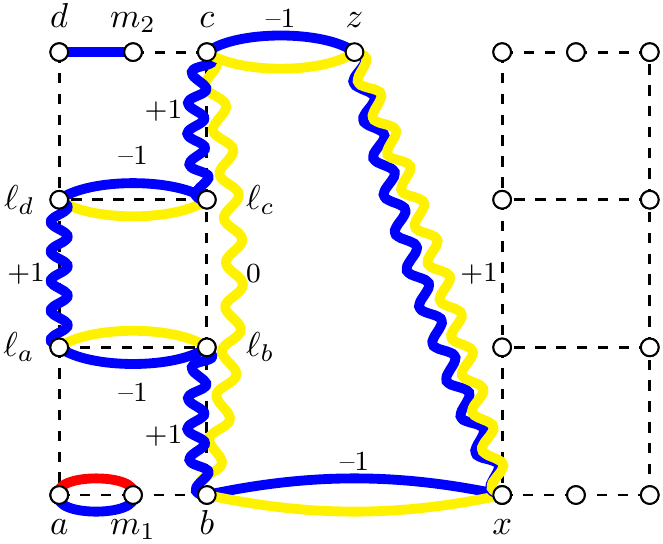}\hspace{2em}\label{fig:cc2}
  }\qquad
  \subfigure[Case 3: $\alpha_2$ contains $m_2c$ and $dz\notin\alpha_1$.]{
    \hspace{2em}\includegraphics[scale=0.7]{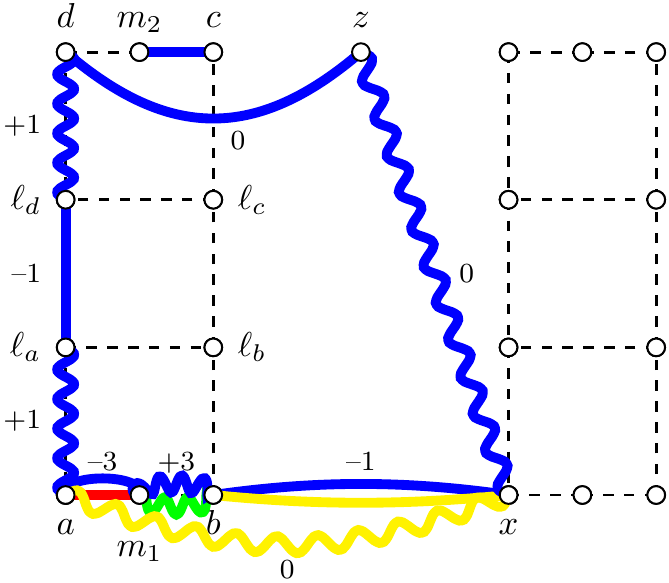}\hspace{2em}\label{fig:cc3}
  }\qquad
  \subfigure[Case 4: $\alpha_2$ contains $m_2c$ and $dz\in\alpha_1\cap\alpha_2$.]{
    \hspace{2em}\includegraphics[scale=0.7]{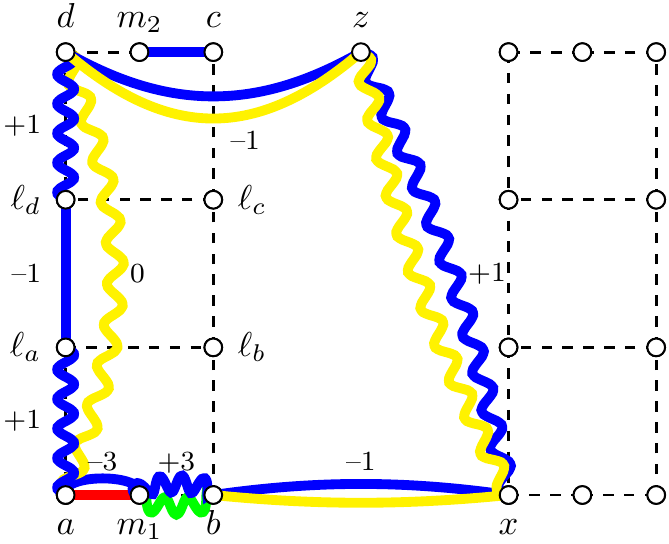}\hspace{2em}\label{fig:cc4}
  }\qquad
  \caption{Different cases that arise when disposing of unsupported common $C$--$C$-edges.
  The dashed edges represent the underlying edge gadgets; adjacencies of $\alpha_2$ are blue,
  adjacencies of $\alpha_1$ are yellow, red, and green. Wavy lines are the new suggested
  adjacencies that should be exchanged for the present ones in the non-negative cycles.} \label{fig:cc}
\end{figure*}

\begin{prop}\label{p:cc}
We may suppose that there are no common $C$-edges other than port edges.
\end{prop}
\begin{proof}
Let $xb$ be a common $C$--$C$-edge in $\alpha_1\cap\alpha_2$.
In the proof, we will refer to and use the notation of Fig.~\ref{fig:cc}.
From what we have proved so far, we may assume that $\alpha_1$ contains the rung edges $\ell_a\ell_b$ and $\ell_c\ell_d$
(Proposition~\ref{p:med}), $am_1$ is a common adjacency of $\alpha_1$ and $\alpha_2$, and either $m_2c$ or $m_2d$
is included in $\alpha_2$ (Proposition~\ref{p:nom}).

First, assume the latter case that $m_2d\in\alpha_2$ (Fig.~\ref{fig:cc1} and \ref{fig:cc2}).
Since the $L$-edges are weakly supported, either $\ell_b\ell_c\in\alpha_2$ (Fig.~\ref{fig:cc1})
or both $\ell_a\ell_b$ and $\ell_c\ell_d$ belong to $\alpha_2$ (Fig.~\ref{fig:cc2}). In either case, we can add ladder
edges to form an alternating $b$--$c$-path with score $+1$ that will be a part of our non-negative
pair of cycles.

Let $cz$ be an adjacency in $\alpha_2$. Since $m_2$ and $\ell_c$ are already matched to different vertices,
$cz$ is unsupported. Now, either $cz\notin\alpha_1$ and $xb\ldots czx$ is a non-negative cycle (see Fig.~\ref{fig:cc1}),
or $cz$ is a common edge and we will also have to exchange some edges in $\alpha_1$. In particular,
$xbczx$ and $xb\ldots czx$ is a non-negative pair of cycles (see Fig.~\ref{fig:cc2}).

Similarly, we can prove the other case when $m_2c\in\alpha_2$; the non-negative cycle pairs
are depicted in Fig.~\ref{fig:cc3} and \ref{fig:cc4}. It can be easily checked that the proof
also works when extremities $x$ and $b$ belong to the same edge gadget (in this case $x$ coincides with $c$
or $d$, and $b$ coincides with $z$). A $C$--$C$-edge connecting two corners of a single port is
ruled out by Proposition~\ref{p:nom}.

Note that if $\alpha_1$ and $\alpha_2$ have a common $z$-edge (Fig.~\ref{fig:cc2} and \ref{fig:cc4}),
we may create a new common unsupported $C$--$C$-edge $xz$. However, the number of common unsupported
$C$--$C$-edges is decreased by 1 in all cases.
\end{proof}

\begin{cor}
We may suppose that all the common adjacencies of the ancestors $\alpha_1$
and $\alpha_2$ are weakly supported: $\alpha_1\cap \alpha_2 \subseteq \Pi$.
More specifically, we may suppose that the only common adjacencies are port
edges and rung edges. Consequently, each unsupported adjacency except for 
rung edges in $\alpha_2$ contributes zero to the score.
\end{cor}

We say that $\alpha_1$ is \emph{uniform} at a vertex gadget, if all the port edges in the gadget
have the same color (they all agree with either the $\pi_1$ edges or the $\pi_2$ edges).
Next, we prove that $\alpha_1$ may be assumed uniform at all gadgets. Such an ancestor $\alpha_1$
directly corresponds to a cut in $G$.

Here, we use the fact that $G$ is cubic: Imagine that $G$ was a complete bipartite graph $K_{n,n}$
with one more vertex connected to all the other vertices. Then our reduction would not work,
since the optimal ancestors would color one bipartition red, the other green, and the extra vertex
half green half red (i.e., half of the ports would be green and the other half red).

First, let us characterize how the non-uniform gadgets look like.

\begin{figure*}[h!]
  \centering
  \subfigure[A non-uniform ancestor $\alpha_1$ at a vertex gadget.]{
    \includegraphics[scale=0.7]{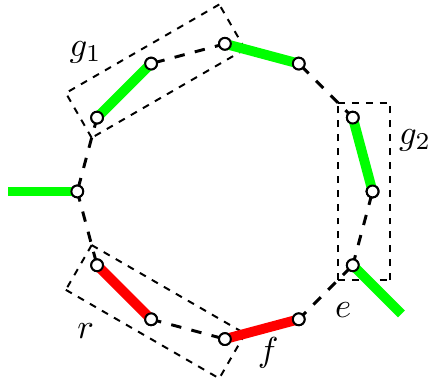}\label{fig:nonunif1}
  }\qquad
  \subfigure[The non-uniform gadgets are connected by unsupported $I$-edges and $C$-edges.]{
    \includegraphics[scale=0.7]{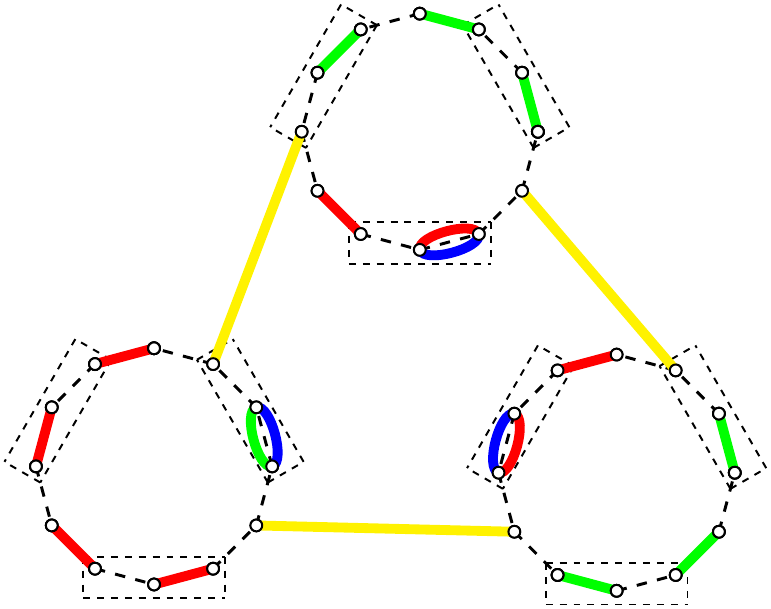}\label{fig:nonunif2}
  }\qquad
  \subfigure[A non-negative cycle used for disposing of non-uniform gadgets and unsupported edges in $\alpha_1$.]{
    \includegraphics[scale=0.7]{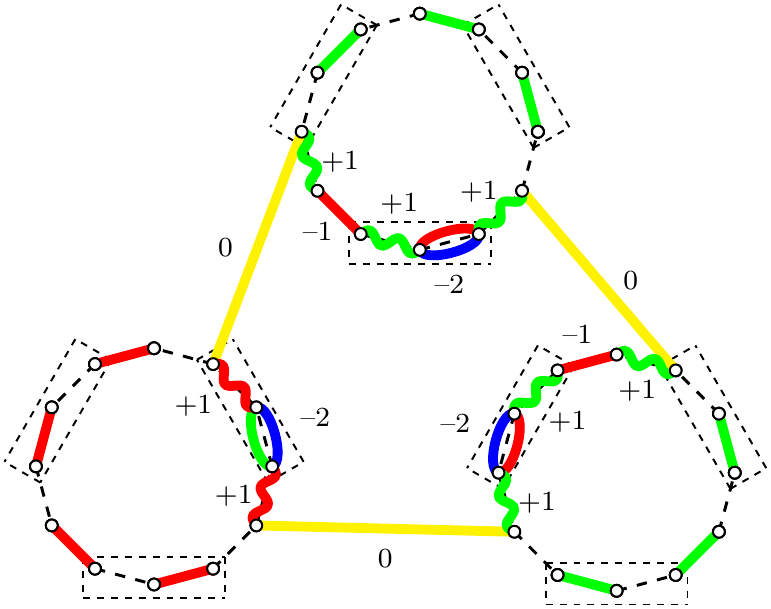}\label{fig:nonunif3}
  }\qquad
  \caption{Non-uniform ancestors at a vertex and a way how to remedy them.} \label{fig:nonunif}
\end{figure*}

\begin{prop}
We may suppose that the following statements are equivalent:
\begin{itemize}
\item $\alpha_1$ is not uniform at a vertex gadget
\item there is one unsupported $I$-edge in $\alpha_1$ incident to the vertex gadget
\item there is one unsupported $C$-edge in $\alpha_1$ incident to the vertex gadget
\end{itemize}
\end{prop}
\begin{proof}
Let $\alpha_1$ be non-uniform at a vertex gadget. Without loss of generality,
let two of the port edges be green and one be red (see Fig.~\ref{fig:nonunif1}).
Denote $r$ the red and $g_1$ and $g_2$ the green edges, such that $g_1$
is closer to $r$ (as in  Fig.~\ref{fig:nonunif1}).
The edge incident to the intermediate extremity between $r$ and $g_1$
is an unsupported $I$-edge.

Obviously, if two \emph{neighbouring} extremities in a vertex gadget are
incident with unsupported edges, there is an augmenting cycle, so we may
suppose that 
the intermediate edge between $g_1$ and $g_2$ is green and one
of the intermediate edges $e$ or $f$ in Fig.~\ref{fig:nonunif1} belongs to $\alpha_1$; the other corner
has an unsupported $C$-edge.

Conversely, if there is an unsupported $I$-edge or $C$-edge,
the neighbouring ports cannot have edges of the same color
(this would imply two neighbouring extremities with unsupported edges in $\alpha_1$).
\end{proof}

We are ready to prove the Normal form lemma.

\begin{prop}
We may suppose that in each vertex gadget, the port edges of $\alpha_1$ are either all red or all green.
Thus, we may suppose that all adjacencies in $\alpha_1$ are supported: $\alpha_1\subseteq\pi_1\cup\pi_2$.
\end{prop}
\begin{proof}
We prove that for each vertex gadget, we may simply look at the three port edges and
choose the color by majority vote. 
In the previous proposition, we have proved that non-uniform gadgets have exactly
two unsupported edges so they form cycles as in Fig.~\ref{fig:nonunif2}.
Fig.~\ref{fig:nonunif3} shows the non-negative cycle that we get by including
the edges decided by majority vote. In each vertex gadget, we may lose 1 point for 
switching the port edge (if this was a common edge), but we get 1 extra point
for increasing the number of supported edges.
\end{proof}

\begin{figure*}[ht]
  \centering
  \includegraphics[scale=0.7]{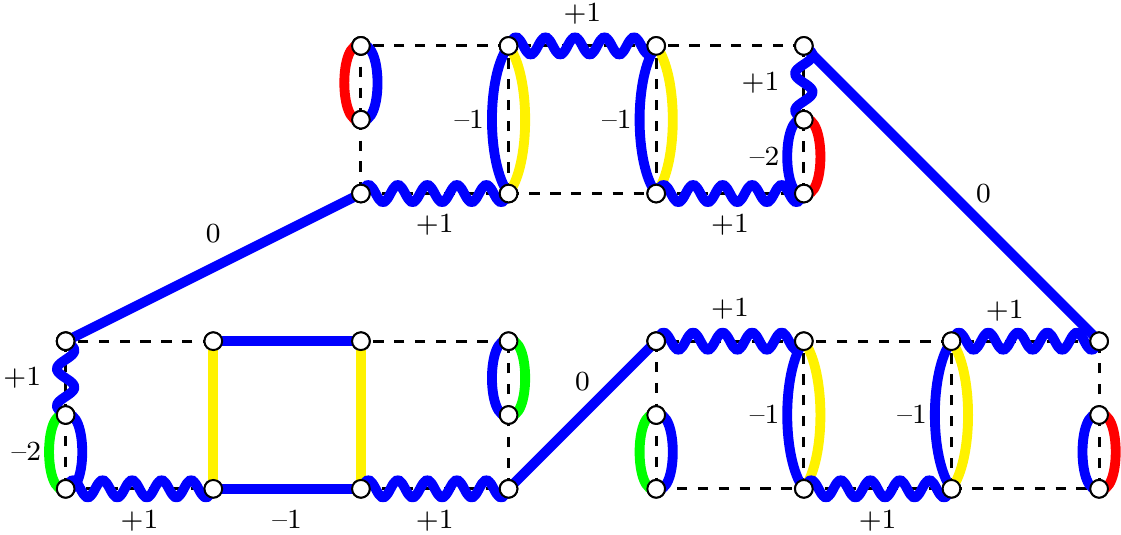}
  \caption{Example of three edge gadgets connected in a cycle by unsupported $C$--$C$-edges.
  We can join two corners with unsupported $C$--$C$-edges in an edge gadget by a non-negative
  path. Note that we also get rid of the blue rung edges in the top and right edge gadgets at the same time.} \label{fig:egcycle}
\end{figure*}

\begin{prop}
We may suppose that all adjacencies in $\alpha_2$ are supported: $\alpha_2\subseteq\pi_3\cup\pi_4$.
\end{prop}
\begin{proof}
The only remaining unsupported edges in $\alpha_2$ are the rung edges and $C$--$C$-edges.
If $\alpha_2$ contains a rung edge, it must in fact contain both rung edges and in the adjacent ports,
only one corner is covered by a port edge. Thus, the edge gadget is incident to two unsupported $C$--$C$-edges.

Conversely, it is easy to see that if $\alpha_2$ contains a $C$--$C$-edge, in the incident edge gadgets,
$\alpha_2$ contains either both rung or both middle rail edges and there are $C$--$C$-edges incident
to the corners of the opposite ports. So the edge gadgets together with the unsupported $C$--$C$-edges
form cycles and all the rung edges are in these edge gadgets (see Fig.~\ref{fig:egcycle}).

In each edge gadget, we can join the two corners by a non-negative alternating path (see Fig.~\ref{fig:egcycle});
we can lose 1 point for destroying a common adjacency of $\alpha_1$ and $\alpha_2$, but we gain 1 point for
increasing the number of supported edges in $\alpha_2$. By exchanging edges along these cycles
we fix both the unsupported $C$--$C$-edges and rung edges.
\end{proof}

This concludes the proof of the Normal form lemma and thus also the proof of NP-hardness and APX-hardness of the \quartet\ problem.

%% file: conclusion.tex
\section{Conclusion}

In this paper, we have settled several open problems concerning
the computational complexity of different rearrangement problems in
the breakpoint models. There are at least three intriguing questions in this
area which remain open. The first two are of theoretical interest and are related
to approximability of the \phylo\ problem, the third question is more
practical:
\begin{enumerate}
\item How well can we approximate \phylo? For example, \quartet\
problem can be easily formulated as an integer linear program (we can
use different variables for the edges present only in $\alpha_1$, only in $\alpha_2$,
and in the intersection $\alpha_1\cap\alpha_2$). Its relaxation might
lead to an algorithm with a good approximation ratio.
\item In the Steinerization approach to ancestral reconstruction,
we repeatedly replace the ancestral genomes by medians of genomes
in the neighboring nodes of the tree until we converge to a local optimum.
Despite the fact that this is the most common approach to ancestral 
reconstruction (also in the other models) and that preliminary
experiments with simulated data suggest that this heuristic performs
very well, no guarantees are known for the method (in any model).
\item Finally, the motivation behind the general breakpoint model
is that we can solve the median problem in polynomial time.
Using the Steinerization method, we can also get very good solutions
of the \phylo\ problem rapidly. The question is: Are these solutions
useful in practice? Are they biologically plausible? Or can we adjust them
and use them as starting points in more complicated models?
\end{enumerate}

%% file: main.bbl
\begin{thebibliography}{23}
\bibitem[{Pe'er and Shamir(1998)}]{peer}
\bibinfo{author}{I.~Pe'er}, \bibinfo{author}{R.~Shamir}, \bibinfo{title}{The
  median problems for breakpoints are NP-complete},
  \bibinfo{journal}{Electronic Colloquium on Computational Complexity (ECCC)}
  \bibinfo{volume}{5}~(\bibinfo{number}{71}).
\end{thebibliography}
